
\magnification\magstep1
\baselineskip 12pt
\parskip=6pt

\tolerance=10000 \hyphenpenalty10000 \exhyphenpenalty10000
\def\pp{\parshape 2 0truecm 15truecm 2truecm 13truecm}
\def\apjref#1;#2;#3;#4; {\par\pp#1, {#2}, {#3}, #4}
\def\bookref#1;#2;#3; {\par\pp#1, { #2}, {\rm #3}}
\def\prepref#1;#2; {\par\pp#1, {#2}}
\overfullrule=0pt
\def\cl{\centerline}

\def\page{\vfill\eject}

\def\ltsima{$\; \buildrel < \over \sim \;$}
\def\lsim{\lower.5ex\hbox{\ltsima}}
\def\gtsima{$\; \buildrel > \over \sim \;$}
\def\gsim{\lower.5ex\hbox{\gtsima}}

{\line{\hfil March 1994}}
\vbox{\vskip 3truecm}
\bigskip
\cl{ \bf EQUATIONS OF GRAVITATIONAL}
\vskip .1cm
\cl{\bf INSTABILITY ARE NON-LOCAL}
\vskip 1.0 truecm
\bigskip\bigskip\bigskip
\vskip 1cm
\bigskip
\cl{\bf Lev Kofman}
\bigskip
\cl{Institute for Astronomy, University of Hawaii, Honolulu, HI
96822}
\bigskip\bigskip
\cl{and}
\bigskip\bigskip
\bigskip
\cl{\bf Dmitry Pogosyan}
\bigskip
\cl{ CITA, University of Toronto, Toronto, ON M5S 1A7,Canada}
\bigskip\bigskip\bigskip\bigskip\bigskip
\cl{  Submitted to: {\it  The Astrophysical Journal}}

\vfill\eject
\vglue 3 truecm
\cl{\bf ABSTRACT}

Few recent generations of cosmologists have solved non-local newtonian
equations of the gravitational instability in an expanding universe.
In this approach pancaking is the predominant form of first collapsing objects.
 Relativistic
counterparts of these equations contain the electric and magnetic parts of the
Weyl tensor. In the linear theory the magnetic part is associated with
gravitational waves. If the magnetic part is ignored, then the newtonian limit
of the relativistic equations is reduced  to the closed set of the local
Lagrangian equations. Recently this fact drew much attention since the
gravitational instability in that form would greatly simplify the study of
cosmic structure formation. In particular, the filamentary structure of
collapsing is predicted. In this paper we resolve the contradiction between the
newtonian theory and relativistic version adopted in some recent papers. We
show that dropping the magnetic part from the basic relativistic equations is
{\it incorrect}. The correct newtonian limit is derived by the  $1/c$-expansion
of the GR equations and the Bianchi identities for the Weyl tensor. The last
ones begin with $\sim 1/c^3$ order, therefore one {\it must} take into account
the  magnetic part in the post newtonian order $\sim 1/c^3$, which contains
non-local terms, related to the non-local gravitational interaction.
For the first time we rigorously show that the basic GR equations with the
magnetic part are reduced {\it precisely} to the canonic newtonian
{\it non-local} equations.  Thus, the correct treatment of the relativistic
version of the gravitational instability resurrects the canonic  picture of the
structure formation.

{\it Subject headings:}\ cosmology: theory  --- large-scale structure
of the universe

\page
\centerline{\bf 1. INTRODUCTION}
\medskip
The cosmological gravitational potential related to the
large scale structure within the horizon is small in the sense of
General
 Relativity,
$\phi \ll c^2$, where $c$ is the speed of light.
Motion of matter is also relatively slow, $v \ll c$. Hence within the
cosmological horizon one can use the Newtonian gravity to describe
the gravitational instability of matter in an expanding universe
(Zel'dovich and Novikov 1983, Peebles 1980).
Newtonian theory is also applicable in the most of cases of
gravitationally
maintained astronomical systems ( Binney and Tremaine 1987).
 Newtonian theory formally
can be derived from the $1/c$ decomposition of the Einstein equations
in the limit $c \to \infty$. The Poisson equation, linear
 with respect to $\phi$,
arises in $1/c^2$ order of the GR equations for gravitational field.

The set of  equations describing the self-consistent evolution of the
pressureless matter of density
$\rho$
with newtonian gravity is non-linear in terms of $\rho, \vec v$ and $
\phi$.
It is clear, intuitively, that the instability of the media due to
the
gravity in the newtonian theory is also non-local, since the
gravitational potential $\phi$ at a given point is a superposition
of contribution over all distributed masses, as textbooks teach us
(Zel'dovich and Novikov 1983, Peebles 1980, Tremaine and Binney
1985).
The dynamics  of the non-linear non-local
gravitational instability is very complicated even in the reliable
newtonian
 theory, and requires the N-body simulations of the non-local
gravitational
 interactions and motions of matter.
Therefore any analytical advantage would be greatly appreciated.
 There are very few successful analytical
approximations to the solution of the basic equations.
One of the best known in the cosmological context
 is the generic Zel'dovich approximation
 (ZA, Zel'dovich 1970).
ZA describes the displacement of particles with the initial velocity
 in the quasi-linear regime, which
 is realized at early stages of non-linear evolution, or at
sufficiently
large scales, where evolution is also quasilinear. ZA is local, in a
sense
that particle evolution upon gravitational forces truncated
along  ZA,
 depends on its Lagrangian position only.
ZA is an approximation and fails after the particle orbits' crossing.
ZA quantitatively predicts that the first non-linear objects in any
cosmological
 scenario
are two-dimensional pancakes, which then evolve into a cellular
structure
with the most mass in the clumps or filaments, depending on the
spectra. Further, non-linear evolution looks like
complicated superposition of hierarchical clustering and pancaking
over a vast
range of scales, with qualitative characteristics crucially depending
 on the model (see, e.g., Shandarin \& Zel'dovich 1989,
 Kofman et al 1992 for theoretical outlook,
Cen \& Ostriker 1992, Bertschinger \& Gelb 1991 for numerical
outlook).

One would expect that relativistic generalization of the
self-consistent
matter evolution in an expanding universe is much more complicated
because of the additional non-linearities in the Einstein equations.
In 1971 Ellis reviewed an elegant form of  GR equations
together with the Bianchi identities of the self-consistent
matter evolution, suggested earlier by Tr\"umper 1967.
 In GR, one can split the Riemann tensor,  $R_{iklm}$,
 which entirely describes the
structure of gravitational field, into the Ricci tensor $R_{ik}$
defined by the Einstein equations, and the Weyl tensor $C_{iklm}$
(e.g., Landau \& Lifshitz 1980).  One can rewrite the Bianchi
identities in the form  rather reminding of the Maxwell equations
(e.g., Hawking and Ellis 1973).
Additionally, one can
split the Weyl tensor into the ``electric'' part $E_{ij}$, and the
``magnetic'' part $H_{ij}$, due to  some of  their similarity to
 electrodynamical counterparts.

 In general relativity
$00$ component of the Einstein Eqs. -- relativistic
counterpart of the Poisson equation --
 plays the role of the constraint
 equation.  The constraint equation
 is automatically resolved for an arbitrary time moment
$t > t_0$  if  it is resolved
 once at the initial
hypersurface $t=t_0$, and the  evolution  equations are satisfied.
Note that
  in the Newtonian limit the  Poisson
equation has to be resolved at each time step $t$.
 Ellis gave a set of evolution equations
for  $E_{ij}$, and  $H_{ij}$. Equation for  the Lagrangian derivative
of
 $E_{ij}$ contain their
 algebraic combination, except one non-local term -- the space
derivatives
of the magnetic part  $H_{ij; k}$.

For the electric part in the $c \to \infty$ limit we put
$E_{ij} ={1 \over c^2}E^{(N)}_{ij}+  {1 \over c^3}E^{(PN)}_{ij}
+O{\left({1 \over c^4}\right)}$.
 The Newtonian limit of the electric part is reduced to the tidal
forces
$$
E^{(N)}_{\alpha \beta}=\nabla_{x_{\alpha}} \nabla_{x_{\beta}} \phi
 - {1 \over 3}\delta_{\alpha \beta}\nabla^2_x \phi ,\eqno(1)
$$
$E^{(PN)}_{ij}$ is the first post-newtonian term.
Respectively, for the magnetic part
$$
H_{ij} ={1 \over c^2}	H^{(N)}_{ij}+
  {1 \over c^3} H^{(PN)}_{ij}+ O{\left({1 \over c^4}\right)}
,\eqno(2)
$$
The magnetic part in the linear theory contains contributions
from
the vector and tensor modes only. Vector modes are irrelevant for
cosmological
 gravitational
 instability,  contribution from  cosmological gravitational waves
is negligible,
they are generated by matter in $1/c^5$  order (in the Lagrangian) only
 (e.g. Landau \& Lifshitz 1980).
 Therefore we can put the newtonian term
$H^{(N)}_{ij}=0$, and $ H^{(PN)}_{ij} \not =0 $ is the first non-vanishing
post-newtonian term. There are controversies in the literature
at this point. For instance, it is incorrect to say that
 the magnetic part is non-zero in the newtonian limit in the absence
of
 vector and tensor modes. It is non-zero in the first
 post-newtonian approximation
only.

Since the magnetic part is vanishing in the Newtonian limit, there is
a temptation to put it always equal to zero   in the evolution
equations (Barnes \& Rowlingson 1989, Mataresse et all 1993).
Then one could obtain the closed set of the Lagrangian {\it local}
(but non-linear) equations
for $E_{ij}, \vec v$ and $\rho$. It looks surprisingly   simpler than
the non-local equations of the Newtonian theory, since one can deal
with the
set of ordinary differential equations.

Recently  this  fact drew much attention
in connection with possible application to the  dynamics of
gravitational
instability
beyond the linear theory
(Mataresse et all 1993, Croudace et al 1994, Bertschinger \& Jain
1994).
Some far-reaching conclusions were made based on this interpretation
 of the relativistic equations.  For the closed set of  the
Lagrangian
 equations which essentially are the
 ordinary differential equations,  the
phase diagram can be drawn. The instability of
 collapse in this theory was found  (Mataresse et all 1993).
Bertschinger \& Jain 1994 concluded that typical collapsed
configuration
are strongly prolated (one dimensional)
 filaments rather than Zel'dovich pancakes.

The implication of these new equations for the Newtonian theory
remained unclear.  There is a clear contradiction with the canonic
newtonian picture of the gravitational instability in cosmology.
 Bertschinger \& Jain 1994 cautiously noted the lack of any
Newtonian derivation of the set of the closed Lagrangian equations
 in this approach. However, the confrontation of views has not been
resolved.
In this paper we give the resolution of this contradiction.
In Sec 2 we recall the basic equations in the Newtonian theory,
and prove that they are non-local for physically interested
 values, like the tidal forces.
 We derive the
newtonian equations for the tidal force (1). It will be
 a convenient form  in order to compare them later with the
relativistic calculations.
In Sec 3 we rigorously implement the transition to  the newtonian
limit of the relativistic equations reviewed by Ellis 1971.
We show that the equation for  $E_{ik}$ begins with $1/c^3$ order,
therefore  the term with the magnetic part $H_{ij;k}$ involving in
 this equation,
has to be calculated in $1/c^3$ order, i.e. beyond the linear
newtonian limit. To calculate the components of the Riemann tensor up
to
$1/c^3$ terms, we have to take into account the first post-newtonian
corrections in the Einstein equations -- the classical problem,
considered many years ago by Einstein, Infeld and Hoffman (1938).
Using this result, we derive the  equations reviewed by Ellis 1971
 in the first
non-vanishing order of $1/c$ decomposition, which turns out to be
of $/c^3$ order.
We demonstrate, that
the resulting equations are precisely reduced to the results of the
newtonian theory. In the inverse order, from the newtonian equations
we can rederive the GR version of the basic equations  in the first
non-vanishing order of $1/c$ decomposition, and show that
 the term related to the
magnetic part inevitably arises. Thus, dropping the magnetic part
from the basic GR equations would have a dramatic effect
similar to the dropping of the magnetic part from the
Maxwell equations.

Among the published works so far, our view of the problem is
closest to that of
 Matarrese et al. 1994, who also claimed that putting
$H^{(N)}_{ij}=0$
cannot be exact.
They used the second order in the cosmological perturbations series,
 and found  gravitational waves beyond of the horizon, whose nature is not
clear to us.
We use $1/c$-series up to $1/c^3$ order in metric $g_{ik}$,
 where gravitational waves
are not yet generated and do not interact with matter,
and therefore need not be considered. It would be interesting to
perform both
decompositions at once.

 We conclude that the correct
basic equations of gravitational instability are intrinsically
non-local.
In Sec. 4 we additionally discuss the Zel'dovich solution,
which is the local approximations to the cosmological gravitational
 instability. We give the explicit solution for the tidal forces in
the ZA.
In the conclusive Section 5 we argue that two-dimensional pancaking
is the generic form of collapsing in the single stream regime.

\vskip 1 cm
\centerline{\bf 2. NEWTONIAN TIDAL FORCES ARE NON-LOCAL}
\bigskip
{\sl 2.1 Newtonian equations via potentials}

\vskip .2 cm

In the epoch of the LSS formation, most of the mass is in the
 form of dark matter
of relic origin (for instance,  the Cold Dark Matter), with no
pressure;
nonlinear Large Scale Structure is
 originated by gravitational instability of small
initial fluctuations.
Let $\vec x,  \vec v =a{d\vec x \over d t},  \rho  (t, \vec x)$ and
$\phi (t,\vec x)$ be, respectively, the comoving coordinates,
peculiar velocity
and density of dark matter (neglecting baryons), and peculiar
Newtonian
gravitational potential. In the Newtonian theory, the motion of  dark
matter
 before the
particle orbits cross obeys a nonlinear system of equations
$$
{\partial \rho \over \partial t} +3 {\dot a \over a} \rho +
 {1\over a}\nabla_x(\rho \vec v)=0,
\eqno(3)
$$
$$
{\partial \vec v \over \partial t} + {1 \over a}(\vec v \cdot
\nabla_x )\vec v
+ {\dot a \over a} \vec v = -{1 \over a} \nabla_x \phi, \eqno(4)
$$
$$
{\nabla_x}^2\phi = 4\pi G a^2 (\rho - \bar \rho),    \eqno(5)
$$
where $a(t)$ is a scalar factor of an expanding Universe and
$\bar \rho$ is a mean density.
These equations are valid in the single stream regime, and admit the
evident
generalization in the regions of multiple streams. \par

For the sake of  simplicity, we  assume that
inflation produces a flat universe $\Omega= 1$, and $\Lambda=0$.
The growing mode of adiabatic perturbations  $D(t)$
in the
Einstein-de Sitter universe is
$D(t)=a(t)  \propto t^{2/3}$.  However, all results reported here are
valid for an arbitrary background cosmology.
The solution of the Poisson equation (5) is non-local, therefore
the motion of the fluid element upon the gravitational forces in the
 right hand side
of the Euler equation (4) is also non-local, i.e is not entirely
defined by
 the masses along the particle trajectory.

One can also conclude that the derivatives of the gravitational
force -- the tidal
 force -- given, for instance, by equation (1), are also non-local.
However it would be useful to write down non-local equations for the
newtonian
tidal forces, or some other
combinations of its further derivatives (Kofman 1993),
especially since  there are opposite statements in the literature
(e.g. Lachieze-rey 1993).
For this purpose let us perform  further analysis of the basic
equations
 (3)-(5).
It is convenient to use the growing solution  $D(t)=a(t)$ as a new
time
 variable instead
 of $t$ (Gurbatov et al 1989, Kofman 1991),
 and introduce a comoving
velocity $\vec u ={d\vec x \over da}=\vec v /{a \dot a}$ in
 respect with this time
variable.
  Then eq.~(4) acquires the form
$$
{\partial  \vec u \over \partial a}+(\vec u \cdot \nabla_x)\vec u
 = (a\dot a)^{-2}\biggl( (3a^2 \ddot a) \vec u - \nabla_x\phi \biggr)
\eqno(6)
$$
 Let $\Phi$ be
 velocity potential so $\vec u = \nabla_x\Phi$.
  We introduce the combination
$A=({3 \over 2}H^2a^3)^{-1}= -(3 \ddot a a^2)^{-1}=(4\pi G\bar
\rho_0)^{-1}$,
 which does
 not depend on time in the
matter-dominated
 Einstein-de Sitter Universe.
 Then from eq.~(6)
 we find
a general relation between the velocity potential and the Newtonian
gravitational potential
(Kofman and Shandarin 1988):
$$
{\partial \Phi \over \partial a}+{1\over2}(\nabla_x\Phi)^2=
 -{3 \over {2a}}  \biggl( \Phi +A \phi \biggr).
\eqno(7)
$$
This equation is often referred to  as the Bernoully equation.
This form is very convenient to discuss the Zel'dovich approximation,
in which the right hand side is vanishing (Kofman 1991).
Substituting density $\rho$ from the right hand side of eq.~(5)
 into eq.~(3), we  find the second equation linking
scalar potentials $\phi$ and $\Phi$
$$
\nabla_{x_{\alpha}} \biggl[{\partial  \over \partial a}(a
\nabla_{x^{\alpha}}
\phi)+ (A^{-1}+a\nabla_x^2\phi)\nabla_{x^{\alpha}}\Phi \biggr] =0.
\eqno(8)
$$
 Thus we obtain two  equations (7), (8)
for two scalar fields $\Phi$ and $\phi$ in the single stream regime.
Note, that from (7) one can express $\phi$ via $\Phi$ and its
derivatives, and
substitute
it in this form into (8). Then one gets the single non-linear partial
differential equation for $\Phi$ only.

\vskip .5 cm
{\sl 2.2. Tidal forces in Newtonian theory}
\vskip .2 cm

The structure of  equation (8) is $\nabla_x \cdot \vec \Sigma=0$,
where
the vector $\vec \Sigma (\vec x) $ is the expression in the square
 brackets in
 eq.~(8). Using the Helmholtz theorem of the vector calculus, we find
$$
\vec \Sigma (\vec x) = \nabla_x \times \biggl[ {1 \over 4\pi}
 \int {d^3x' \over
{|\vec x - \vec x'|}}  \nabla_{x'} \times \vec \Sigma (\vec x')
\biggr].
\eqno(9)
$$
Substituting  the expression for  $\vec \Sigma (\vec x) $
from the square bracket of eq. (8), in eq. (9),
after some vector algebra, we end up with the formula
$$
{\partial  \over \partial a}(a \nabla_{x^{\alpha}} \phi)+
A^{-1} \nabla_{x^{\alpha}} \Phi = { a \over 4\pi}
 \nabla_{x^{\beta}} \nabla_{x^{\alpha}}  \biggl[  \int {d^3x' \over
{|\vec x - \vec x'|}} (\nabla_{x'_{\beta}} \Phi) (\nabla_{x'}^2 \phi)
\biggr].
\eqno(10)
$$

In the Newtonian theory we can construct the traceless symmetric
tensors
$$
\sigma^{(N)}_{\alpha \beta}={1 \over 2}(\nabla_x^{\beta} u_{\alpha}+
\nabla_x^{\alpha}u_{\beta})-
 {1 \over 3}\delta_{\alpha \beta}\nabla_x^{\gamma} u^{\gamma}=
\left(\nabla_{x^{\alpha}} \nabla_{x^{\beta}}\Phi -
 {1 \over 3}\delta_{\alpha \beta}\nabla^2_x \Phi\right),
$$
along with the traceless tidal forces tensor (1).
 Taking derivatives
$\nabla_{x^{\alpha}} \nabla_{x^{\beta}}$ from eq.~(7), we  can derive
the  equation for  the Lagrangian time derivative of
$\sigma^{(N)}_{\alpha \beta}$ along the trajectory
$$
{ D \over D a}   \sigma^{(N)}_{\alpha \beta}=-{3 \over 2a}
\left( \sigma^{(N)}_{\alpha \beta} +A E^{(N)}_{\alpha \beta} \right).
\eqno(11)
$$
Taking derivative $\nabla_{x^{\beta}}$ of eq.~(10), we derive the
equation for the evolution of the tidal forces along of trajectory
$$
{ D \over D a}   E^{(N)}_{\alpha \beta} +
{1 \over a} \left( E^{(N)}_{\alpha \beta} +
A^{-1} \sigma^{(N)}_{\alpha \beta} \right) -
(\nabla_{x^{\gamma}}\Phi)(\nabla_{x^{\alpha}} \nabla_{x^{\beta}}
\nabla_{x_{\gamma}}\phi)
- {1 \over 3} (\nabla^2\phi)(\nabla^2\Phi) \delta _{\alpha \beta} =
$$
$$
={1 \over 4 \pi} \nabla_{x^{\beta}} \nabla_{x^{\gamma}}
\nabla_{x^{\alpha}}
\biggl[  \int {d^3x' \over
{|\vec x - \vec x'|}} (\nabla_{x'_{\gamma}} \Phi) (\nabla_{x'}^2
\phi) \biggr]
{}.
\eqno(12)
$$

At this point we rewrite eq. (12) back in the coordinate time and
physical
variables to compare it with the result we will derive in Sec.3 from
the
relativistic equations:
$$
{\partial \over \partial t} E^{(N)}_{\alpha \beta}
+ {\dot a \over a} E^{(N)}_{\alpha \beta}
 - {4 \pi G  \over 3} a \nabla _{x} (\rho \vec v) \delta _{\alpha
\beta} =
a G \nabla_{x^{\beta}} \nabla_{x^{\alpha}}  \biggl[  \int {d^3x'
\over
{|\vec x - \vec x'|}} \nabla _{x'} (\rho \vec v ) \biggr] .
\eqno(13)
$$
 Note the
non-local integral term selected to
 the right hand side. This non-local term
cannot be reduced to the combination of the local terms.
It prevents us from  obtaining a closed set of Lagrangian equations
in the rigorous Newtonian theory.

\vskip 1 cm
\centerline{\bf 3. TIDAL FORCES IN GENERAL RELATIVITY ARE NON-LOCAL}
\vskip .5 cm
{\sl 3.1. Notations and GR equations}
\vskip 0.2 cm
In General Relativity the gravitational field, described by the
Riemann
tensor $ R_{iklm} $, is  determined by the Einstein equations
$$
{R_i}^k  = {8 \pi G \over c^4 } \left( {T_i}^k - {1 \over 2}
\delta _i^k {T_m}^m \right) \eqno(14)
$$
and the Bianchi identities
$$
2 {C^{iklm}}_{;m} = R^{li;k} - R^{lk;i} +
	 {1 \over 6} \left( g^{lk} R^{;i} - g^{li} R^{;k} \right)
\eqno(15)
$$
Here $ {R_i}^k = {R^{mi}}_{mk}$ is the Ricci tensor, $R = {R_m}^m $
is a curvature, $ C_{iklm} $ is the
Weyl tensor and $()^;$ is a covariant derivative.
Latin letters run from  $0$ to $3$ and Greek letters
run from  $1$ to $3$. The signature of metrics we use is $ (-+++) $.
For a  pressureless matter treated as the perfect fluid,
 energy-momentum tensor is given by
$T^{ik}=\varepsilon u^i u^k $, $\varepsilon $ is the energy density
and $ u^i $
is four-velocity of the fluid, $ u^i u_i = -1 $. Equations of motion
of the
matter contained in  the Einstein equations are
$$
{{T_k}^{k}}_{;i} = 0 \eqno(16)
$$

One can further decompose the Weyl tensor into the "electric" part
 $ E_{ik} = C_{ilkm}u^l u^m $ and the
magnetic $ H_{ik} = {1 \over 2} {\eta _{il}}^{mp} C_{mpkq} u^l u^q $
(where
 $ \eta _{iklm} $ is fully antisymmetric four-tensor):
$$
C_{iklm} = (\eta _{ikpq} \eta _{lmrs} + g_{ikpq}g_{lmrs})u^p u^r
E^{qs}
- (\eta _{ikpq} g_{lmrs} + g_{ikpq} \eta _{lmrs})u^p u^r H^{qs} ,
$$
where
$$
g_{iklm} = g_{il}g_{km} - g_{im}g_{kl}. \eqno(17)
$$
Then the Bianchi identities lead to the ``Maxwell-like''
equations on $ E_{ik} $ and $ H_{ik} $ (Tr{\"u}mper 1967).
 From them of the most interest is the evolution equation on $ E_{ik}
$
which we write down for the case of pressureless irrotational fluid
$$
{p_i}^n {p_k}^l u^m E_{nl;m} + p_{ik} \sigma ^{lm} E_{lm} + \Theta
E_{ik}
- 3 E_{l(i} {\sigma _{k)}}^l
+ p_{l(i} {\eta _{k)r}}^{sm} u^r {H^l}_{s;m}
= -{4 \pi G \over c^4 } \varepsilon \sigma _{ik}. \eqno(18)
$$
Here $ p^{ik} = g^{ik} + u^i u^k $ is the projection tensor.
The velocity derivative tensor
is presented as the sum of its trace $\Theta$, symmetric traceless
shear $ \sigma _{ik} $ and antisymmetric vorticity $ \omega _{ik} $
 as
$ p_{il} {p_k}^m {u^l}_{;m} = \sigma _{ik} + {1 \over 3} \Theta
h_{ik} +
\omega _{ik} $.
In our case  $ \omega _{ik} =0$.
 Parentheses on the pair of indexes denote symmetrization,
while brackets later in the text -- antisymmetrization.
The first term in (18) is the Lagrangian derivative
of the electric part along of the trajectory, other terms are algebraic
 combinations of the local tensors, except one non-local term, containing the
 space derivatives of the magnetic part $ {H^l}_{s;m}$.
The rest of the equations that follow from the Bianchi identity
the reader can find in Ellis 1971. Note that there are two local equations for
$\sigma_{ik}$ and $\Theta$  in the absence of vector and
 tensor modes.

\vskip .5 cm
{\sl 3.2. ${1 \over c}$ expansion of the Weyl tensor for the
 gravitational instability in an expanding universe}
\vskip 0.2 cm

Now we shall analyse the newtonian limit for the GR equations
that describe the motion of the
self-gravitating pressureless irrotational fluid in an expanding
Universe and,
 specifically,
the role of eq.(18) for the electric part of the Weyl tensor in this
limit.

To describe the gravitational instability on the uniform
expanding cosmological background, let us write the metrics $ g_{ik}
$ as
$$
ds^2 = -a^2(\tau) \left[ c^2(1+h_{00}) d\tau^2
 -2h_{0 \alpha}cd\tau dx^{\alpha}-
(\delta_{\alpha \beta} - h_{\alpha \beta}) dx^{\alpha} dx^{\beta}
\right],
\eqno(19)
$$
$\tau$ is the conformal time, for the sake of simplicity considering
 the flat space geometry.

Newtonian  limit of the GR equations formally can be obtained by
taking the limit  $ 1/c \to 0 $. We will need the decomposition of the metrics
$h_{ij}$ up to first post-newtonian correction $\sim 1/c^3$.
The quantities describing matter behave themselves
 as $ \varepsilon \to \rho c^2 $,
$ u^0 \to a^{-1} $, $u_0 \to -a $ and $ au^{\alpha} = c^{-1}
v^{\alpha} $,
where
$\rho $ is the mass density and $ v^{\alpha}$ - physical velocity.
Hence the right-hand side of eq.(14) is of order of $ 1/c^2 $ and, as
we are
interested in self-consistent gravitational field produced by matter,
the
expansion
of $ h_{ik} $ also starts from $ 1/c^2 $ terms. Substitution of the
metric in
form (19) shows that when $ 1/c \to 0 $ only $ h_{00} $ remains
present in
the equations of motion (16):
$$
\eqalignno{
c \left[ \rho^\prime  + 3 {a^{\prime} \over a} \rho +
{\left( \rho v^{\alpha} \right)}_{,\alpha} \right] &= 0 &(20)\cr
{1 \over c^2 } \left[ v^{\alpha \prime} + {a^{\prime} \over a}
v^{\alpha} +
v^{\beta} {v^{\alpha}}_{,\beta} \right] &=
-{1\over 2} a^2 {h_{00}}^{,\alpha} &(21)
}
$$
Here the prime denotes derivative with respect to conformal time $
\tau $ while
ordinary spatial derivatives are denoted by $( )_{, \alpha}$.
The correspondence with standard 3D notation we used in previous Section
 is: $ {\partial \over
\partial
t} = {1 \over a} {\partial \over \partial \tau}={1 \over a}
()^{\prime} $,
$ \nabla _{x^{\alpha}} =
{\partial \over \partial x^{\alpha} } = a^2 {\partial \over \partial
x_{\alpha}} = ()_{,\alpha}$.
 Together with (00) component of the Einstein equations
(14) which determine $ h_{00} $:
$$
{1 \over 2} a^2 {{h_{00}}_{,\gamma}}^{,\gamma}
= { 4 \pi G \over c^2 } a^2 (\rho - \bar \rho) \eqno(22)
$$
the equations of motion, thus, form a closed set
of equations for $h_{00}$, $\rho $
and $ v^{\alpha} $,
identical to the Newtonian ones (3)-(5) if one identify $h_{00}$ with
gravitational potential $ h_{00} = 2 \phi / c^2 $. Next post-newtonian
 correction in $h_{00}$ is $1/c^4$-order.

The remaining components of the Einstein equations simply
determine $ h _{0 \alpha} $ and $ h_{\alpha \beta} $ that are
consistent with the Newtonian limit through the known quantities $
\phi , \rho
,
v^{\alpha}$.
Using well-known formulas
for post-newtonian metric decomposition,
 e.g.  from  Landau \& Lifshitz (1980, \S 106), and generalizing them
including the expansion of the universe,
we have
$$
\eqalignno{
h_{\alpha \beta} &= h_{00} \delta_{\alpha \beta} + O \left({1 \over
c^4}\right)
= {1 \over c^2} 2 \phi \delta _{\alpha \beta} + O \left( {1 \over
c^4} \right),
&(23) \cr
h_{0 \alpha}
&= {1 \over c^3} {1 \over 4 \pi} \int {d^3x' \over
{|\vec x - \vec x'|}} \left( 16 \pi G \rho v_{\alpha} + {1 \over a^4}
{(a^4 \phi _{,\alpha})}^{\prime} \right) . &(24)
}
$$

In the last expression the auxiliary condition  $ a^2 {h_{0
\alpha}}^{,\alpha}
=
{1 \over 2} {(a^2 {h_{\alpha}}^{\alpha} )}^{\prime}$ is assumed.
Let us note that the quantity $ h_{0 \alpha} $ is already non-local.

The expressions (23)-(24) are sufficient to write down the Weyl tensor
$
C_{iklm}$
up to the order $ 1 / c^3 $ and to determine the first non-vanishing
terms for
$ E_{ik} $ and $ H_{ik} $:
$$
\eqalignno{
C_{0 \alpha 0 \beta} &= a^2 E_{\alpha \beta}
&\left({1 \over c^2}\right) ~~~~(25)\cr
C_{0 \alpha \beta \gamma} &=
{a^2 } \left[ h_{0 [\beta ,\gamma] \alpha}
+ {1 \over 2} \left( {h_{0 [\gamma ,\delta}}^{,\delta} -
{h_{0 \delta ,[\gamma }}^{,\delta} \right) g_{\beta ] \alpha} \right]
&\left({1 \over c^3} \right) ~~~~(26)\cr
C_{\alpha \beta \gamma \delta} &= E_{\alpha \gamma} g_{\beta \delta} +
E_{\beta \delta} g_{\alpha \gamma} - E_{\alpha \delta} g_{\beta \gamma}
- E_{\beta \gamma} g_{\alpha \delta}
&\left({1 \over c^2} \right)~~~~(27)\cr
E_{\alpha \beta} &= {1 \over c^2} \left( \phi _{,\alpha ,\beta }
- (1 / 3) g _{\alpha \beta} {\phi _{,\gamma}}^{,\gamma}
\right) = {1 \over c^2} E^{(N)}_{\alpha \beta}
&\left({1 \over c^2} \right) ~~~~(28)\cr
H_{\alpha \beta} &= {\eta _{\alpha \delta}}^{\gamma 0}
C_{\gamma 0 \beta 0} u^{\delta} +
{1 \over 2} {\eta _{\alpha 0}}^{\gamma \delta} C_{\gamma \delta \beta m} u^m
&\left( 1 \over c^3 \right). ~~~~(29)
}
$$
The remaining ($00$) and ($0 \alpha $) components of $E_{ik}$ and
$H_{ik}$
tensors are of a higher order in $ 1/c $.

	Although Newtonian equations (20)-(22) arise as $ \sim 1 /
c^2 $ terms
in the expansion of relativistic equations,
  (and to obtain them one does not need to consider
any higher order quantities), it is no surprise, however, that one can
construct such combinations of GR equations which do not contain
$1 / c^2 $ terms at all.  In this case, to reproduce correct limit $
1/c \to 0$
of such combinations
it is necessary to carry out the calculations beyond $ 1/ c^2 $
order.
 At the first
glance one could think that the new relations between Newtonian
quantities
could arise as a result, but,
of course, it is evident that all these new equations must appear
to be some combination of basic set (20)-(22) in order for GR to be
compatible
with the Newtonian theory.
The evolution equation (18) for $ E_{ik} $ just provides us one such
example.
Let us fix free indexes in eq.(18) to be spatial ones $(\alpha
\beta)$.
Using expressions (28)-(29) and the Newtonian limits for the
shear $ a \sigma _{\alpha \beta} = {1 \over c} \sigma^{(N)}_{\alpha \beta}
$
 and the expansion
  $ a \Theta  = {1 \over c} \Theta ^{(N)}$,
we get in leading order
$$
{1 \over c^3} \left[
E^{(N) \prime}_{\alpha \beta} + {a^{\prime} \over a}
E^{(N)}_{\alpha\beta}
+ E^{(N)}_{\alpha \beta ,\gamma} v^{\gamma} +
g_{\alpha \beta} \sigma ^{(N) \gamma \delta} E^{(N)}_{\gamma \delta}
+ \Theta ^{(N)} E^{(N)}_{\alpha \beta}
- 3 E^{(N)}_{\gamma (\alpha} {\sigma ^{(N)}_{\beta )}}^{\gamma}
+ {4 \pi G } \rho \sigma ^{(N)}_{\alpha \beta} \right]
$$
$$
\eqno = - {\eta _{(\beta 0}}^{sm} {H}_{\alpha) s;m} ~~~~(30)
$$
The left hand side of this equation is of order $ 1 / c^3 $. The magnetic
part of
the Weyl
tensor at the right side is non-vanishing in this order, according to (29).
Hence, it does not
drop out
from eq.(30).  Therefore we conclude that despite the fact that $
H_{ik} $ is
of higher
order of smallness than $ E_{ik} $ it
cannot be set to zero in eq.(18) on the basis of considering
Newtonian limit.
As we show later evolution equation for $ E^{(N)}_{ik} $ derived from
GR
remains to be nonlocal.

In the next section we proceed to prove that Bianchi identities
and, consequently, eqs.(18),(30)
are completely  consistent with standard Newtonian equations (3-5).

\vskip 1.5 cm
{\sl 3.3 Reduction of Bianchi identities into Newtonian
equations }
\vskip 0.2 cm
The straightforward way to investigate eq.(30) in the Newtonian limit
is to
insert expressions (28),(29) for $ E_{ik} $ and $ H_{ik} $ into (30). The
following
algebraic computations are rather lengthy and given in the
Appendix.
It is more convenient to deal directly  with
the Bianchi identities utilizing the components of the Weyl
tensor
 given by (25)-(27).
To get rid of $ R_{ik} $ terms we rewrite eq.(15) using the Einstein
 equations as
$$
{C_{iklm}}^{;m} = { 4 \pi G \over c^4 } \left[ \left( T_{li;k} - {1
\over 3}
g_{li} {T^p}_{p;k} \right) - \left( T_{lk;i} - {1 \over 3} g_{lk}
{T^p}_{p;i}
\right) \right]. \eqno(31)
$$
The principal combinations of  indexes $ (ikl) $ are $ (0 \alpha
0) $,
$ (\alpha 0 \beta ) $ and $ (\alpha \beta \gamma) $. In the leading
order
in $ 1 / c $, combination $(0\alpha 0)$  straightforwardly corresponds to the
derivative of the Poisson equation
$$
{E_{\alpha \beta}}^{,\beta} = {8 \pi G \over 3 c^2} \rho _{,\alpha}
{\left( {\phi ^{,\gamma}}_{,\gamma} - {4 \pi G } \rho \right) }
_{,\alpha} = 0. \eqno(32)
$$
The $(\alpha \beta \gamma)$ combination  leads to the equation
for
the time evolution of the magnetic part $H_{ik}$. However,
 in the Newtonian limit
 all terms,
containing $H_{ik}$, appear to be of higher order than the rest  of
the
equation which gives us again
$$
g_{\alpha \gamma} {\left({\phi _{,\delta}}^{,\delta} - {4 \pi G }
\rho \right)}_{,\beta} - g_{\beta \gamma} {\left( {\phi
_{,\delta}}^{,\delta}
- {4 \pi G } \rho \right)}_{,\alpha} = 0. \eqno(33)
$$
The fact that the relativistic evolutionary equation for $H_{ik}$
reduces
in the Newtonian limit to an identity that does not include time
derivative
of $H_{ik}$ suggests that magnetic part of the  Weyl tensor
does not have an independent dynamical meaning in the Newtonian
theory.

The last combination $(\alpha 0 \beta)$ gives rise, in particular, to
the evolution equation for
$ E_{ik} $ and is slightly more complicated. Substituting (26) into (31),
 we get
$$
- {1 \over c} \left[ E_{\alpha \beta}^{\prime}
+ {a^{\prime} \over a} E_{\alpha \beta} \right]
+{ a^2 \over 2 } {{h_{0 [\delta}}^{,\delta}}_{,\beta] \alpha} =
{4 \pi G \over c^3} \left[ \left( {1 \over 3}
\rho^{\prime} + {a^{\prime} \over a}  \rho \right) g_{\alpha \beta} +
{(\rho v_{\beta})} _{,\alpha} \right] .\eqno(34)
$$
Noticing that the antisymmetric term in the squared brackets
 the left hand side is
$ \left[ ~... \right] = c^{-3} \left[ \phi ^{\prime} _{,\alpha \beta}
+{a^{\prime} \over a} \phi _{,\alpha \beta} - {c^3\over 2} R_{0 \beta
,\alpha}
\right]$
we finally have
$$
{\left( g_{\alpha \beta} {\phi _{,\delta}}^{,\delta} \right) }^{\prime}
+ {a^{\prime} \over a} {\phi _{,\delta}}^{,\delta} g_{\alpha \beta}
= {4 \pi G } \left( \rho ^{\prime} + 3 {a^{\prime} \over a} \rho \right)
g_{\alpha \beta}, \eqno(35)
$$
which is essentially
$$
\left[ {\partial \over \partial \tau} + {a^{\prime} \over a} \right]
\left( \Delta \phi - 4 \pi G a^2 (\rho - \bar \rho) \right) = 0 .
\eqno(36)
$$

If we return to eq. (34) and substitute there $ h_{0 \alpha }$ from eq. (24),
we arrive at the form
$$
{\partial \over \partial \tau} E^{(N)}_{\alpha \beta}
+ {a^{\prime} \over a} E^{(N)}_{\alpha \beta} -
{4 \pi G \over 3} g_{\alpha \beta} (\rho v^{\gamma}) _{,\gamma} =
G a^2 {\left[ \int {d^3x' \over {|\vec x - \vec x'|}}
\rho v^{\gamma} \right] } _{,\gamma \alpha \beta} .  \eqno (37)
$$
which is completely identical to eq. (13) derived from
the
 Newtonian
theory, but
only written in the relativistic notations using
 $\tau$ and $g_{\alpha \beta} $.
Thus, we prove that the relativistic equations for the gravitational
 instability of the scalar perturbations in an expanded universe
are reduced to the newtonian equations, which are non-local.

\vskip 1 cm
\centerline{\bf 4. LOCAL APPROXIMATIONS}
\vskip 0.2 cm

As we demonstrated in the previous sections, rigorous Newtonian
equations
of the gravitational instability in an expanded universe are
intrinsically non-local. However, this does not exclude some
useful intermediate {\it approximations} which might be local.
Famous Zel'dovich approximation (Zel'dovich 1970) is an example of
the
approximated     description of the gravitational instability, which
turns out to be a {local} one.

It is convenient to use the  equation of motion  in the  Lagrangian
 description.
In such a case the dynamics is described by the displacement,
 $\vec S(\vec q, t)$,
of each particle from its initial position $\vec q$. Its current
Eulerian
comoving position,
$\vec x$, is then given by
$$\vec x=\vec q+\vec S(\vec q, t).\eqno(38)$$

For 1D case the mildly non-linear dynamics is quite simple. The
reason is
 that the force
exerted by a density perturbation over a given particle
is independent of its distance to  the particle.
 Therefore,
before any shell crossing, the displacement of  each particle depends
on
its Lagrangian position $\vec q$ only. Then
 the displacement field for the growing mode can be factorized
$\vec S(\vec q,t)=\nabla_{\vec q}\Phi(\vec q)\ D(t).$
 In 1D case the ZA is then identical to the exact dynamics.
In relation (2) the displacement factorizes in a spatial function,
$\nabla_{\vec q}\Phi(\vec q)$, which depends on the  initial
conditions,
 and a universal
time dependent growing mode $D(t)$.
Using equations (7), (8) for the potentials $\phi$ and $\Phi$, one
can
rigorously derive  in 1D case that
$$
\phi=-A^{-1} \Phi , \eqno(39)
$$
and  then the dynamical equation (7) for the velocity potential
$\Phi$
 is reduced to the
``shortened'' equation without the right hand side:
$$
{\partial \Phi \over \partial a}+{1\over2}(\nabla_x\Phi)^2=0.
\eqno(40)
$$

The 3D generalization of the  form (38) of the displacement is the
Zel'dovich
approximation (ZA).
The 3D generalization of the equations (39) and (40) is the
Zel'dovich
 approximation expressed via velocity potential $\Phi$. Note the
simple
 solution
of equation (40):
$$
\Phi(a,\vec x)=\Phi_0(\vec q)+{(\vec x -\vec q)^2 \over 2a}=
\Phi_0(\vec q)+{a \over 2}{\vec u_0}^2  . \eqno(41)
$$
The solution (47) describes the deformations of 3D-hypersurface
of the potential $\Phi$, and is valid until
the formation of folds of $\Phi$-hypersurface, which corresponds to
caustics
(pancakes).

In this section we consider the equations for the tidal forces in ZA,
show that they are local, and find its
solution.
Let us
 introduce the
 tensor of the velocity derivatives $W_{\alpha \beta}(\vec x, t)
=-\nabla_{x^{\alpha}} v_{\beta}$
 in the Eulerian space.
 For the potential motion it is reduced to
$W_{\alpha \beta}=-\nabla_{x^{\alpha}}\nabla_{x_{\beta}}\Phi$. Note
also that
$W_{\alpha \beta}=-\sigma^{(Z)}_{\alpha \beta}-{1 \over
3}\delta_{\alpha \beta}
\Theta^{(Z)}$. Here $\Theta^{(Z)}=\nabla_x^2 \Phi$.
Let $\lambda_i$ be its  eigenvalues.
The field of the  $W_{\alpha \beta}(t)$-tensor evolves in time, its
initial
value (in the Lagrangian space)
 coincides with the Lagrangian  deformation tensor
$D_{\alpha \beta}=-\nabla_{q^{\alpha}}\nabla_{q^{\beta}}\Phi_0$,
$\lambda_{0\alpha}$ are its  eigenvalues.
 We  will call the $S_{\alpha \beta}$-tensor
  the Eulerian deformation tensor.
The tensor of tidal forces is the ZA, with help of  is equation (39),
is
$$
 E^{Z}_{\alpha \beta}=-{1 \over A} \left( W_{\alpha \beta}
 -{1\over 3}\delta_{\alpha \beta}W^{\gamma}_{\gamma} \right),
\eqno(42)
$$
Taking space derivatives of the equation (40) and using its
definition,
we derive the equation for the tidal forces in the ZA.
$$
{ D \over D a}   E^{(Z)}_{\alpha \beta} +
{2 \over 3a} \Theta ^{(Z)} E^{(Z)}_{\alpha \beta}
 + E^{(Z)\tau}_{\alpha} \sigma^{(Z)}_{\tau \beta}
-{1 \over 3a} \delta _{ij} E^{(Z)}_{\gamma \tau} \sigma^{(Z)\gamma
\tau}=0
  . \eqno(43)
$$
This equation is simpler than equation (12). Actually, in 1D case
equation (12) reduced to equation (43).

It is easy to construct the solution of equation (43) for the tidal
forces
 in the ZA
Using the definition of $W_{\alpha \beta}$ and its eigenvalues
  $\lambda_{\alpha}(t)$, one can instantly find from (41) how they
 are related to their initial values $\lambda_{0\alpha}$:
$$
\lambda_{\alpha}(t)=
{\lambda_{0{\alpha}} \over({1+D(t)\lambda_{0{\alpha}}})}. \eqno(44)
$$
Then the principal radii of curvature
$R_{\alpha}=\lambda_{\alpha}^{-1}$
linearly increase with $a(t)$:  $R_{\alpha}(t)=R_{0\alpha}+a(t)$.
It means that the $\Phi$-hypersurface obeys the rule of the
evolution, which is
similar to the Huygens principle of geometric optics for the wave
front.

Substituting (44) into (42), we conclude, that the principal axis of
the
tidal forces tensor are not rotating in the single stream ZA,
and its eigenvalues $\mu_i$ are evolving with time as
$$
\mu_{\alpha}(t)=-{2 \over 3}\left(3\lambda_{\alpha}(t)-
(\lambda_1(t)+\lambda_2(t)+\lambda_3(t) \right).
 \eqno(45)
$$
Thus, the form (45) is the solution of equation (43).
In the work in the progress (Bond et al 1994) we are comparing the tidal
forces in the ZA against the actual tidal forces.

Thus, the tidal forces in the ZA are closely related to the
deformation
 of the 3D hypersurface $\Phi(t)$, described by the simple  Huygens
principle.
In the ZA the  evolved fields  can be  entirely
described via   the initial eigenvalues $\lambda_{\alpha0}$-s --
 ZA is a local approximation.
It may be interesting to consider the solution of the exact single
stream
 equation (12) in terms of decomposition around the form (44), which
is
the exact solution in 1D case.

\vskip 1 cm
\centerline{\bf 5. CONCLUSION}
\vskip 0.2 cm

We implement the $1/c$ decomposition of the basic GR equations,
describing
gravitational instability of the pressureless matter with the scalar
initial
perturbations in an expanded universe.
It allows us to derive the newtonian limit of these equations.
 These equations include the
Bianchi identities, which begin from $1/c^3$ order.
Therefore
we have to take into account the Weyl tensor up to this order.
 It requires
the first post-newtonian corrections to the metric $g_{ik}$.
Having borrowed corresponding results from the textbook
(Landau \& Lifshitz 1980), we conclude that the relevant
post-newtonian
corrections are contained in the magnetic part of the Weyl tensor,
meanwhile the relevant electric part is needed in $1/c^2$ order only.
Using this post-newtonian magnetic part,
we then end up with the canonic newtonian equations for the
gravitational
 instability in an expanded universe.
The reader can find the formula for the magnetic part in the
Appendix.

We should warn the reader about the statements in the literature
 on the magnetic part.
In the absence of the vector and tensor modes, it is incorrect
to say that the magnetic part is non-zero
in the newtonian limit $1/c^2$.
It is correct to say that it is zero in this limit,
but it is not zero in the first post-newtonian limit $1/c^3$.

Thus, we can continue to work with the newtonian equations for the
 gravitational instability. The results reported in the
papers exploited the  set of local Lagrangian equations, are
irrelevant for
the general case of cosmological gravitational instability.
However this approach
is useful for the special class of the solutions of Einstein
equations
constrained by the condition $ H_{ik} \equiv 0 $ (Croudace et al,
1993).

Another controversial question aroused from the recent discussion
regards
to the first collapsing forms.
It was known for a long time that the first collapses occur generally
not
 around the density peaks, associated
with the maxima of the sum
 $\rho \propto \lambda_{01}+\lambda_{02}+\lambda_{03}$,
  but rather around the maxima of the leading component of the
deformation
 tensor $ \lambda_{01}$ ( Shandarin \& Klypin 1983). For the high density
peaks
their positions can be very close.
Initially underdense regions also can undergo collapse, since
maximum $  \lambda_{01}$ can be realized in the region where the sum
$\lambda_{01}+\lambda_{02}+\lambda_{03}$ is negative
 (Zel'dovich \& Novikov 1983).

 Another question is the form of the first
 collapsing objects.
In the general case the local density can  be obtained  by the inverse
of the Jacobian of the transformation (38)  between $\vec q$
and $\vec x$, so that
$$\rho   =  {{\bar \rho} \over {\vert {\partial \vec x} / { \partial \vec
q} \vert }}
 ={{\bar \rho}
 \over {\vert(1-S_{1})(1-S_{2})(1-S_{3})\vert}},
\eqno(45)
$$
where $S_{\alpha}$ are the eigenvalues of the deformation tensor
$\nabla_q^{\alpha} S_{\beta}$, $\vec S(\vec q, t)$ is the
displacement
vector.
The form of the collapsing objects depends on  the rate of decreasing
of
the  factors in the  denominator (45). In this case the leading
eigenvalue $S_1$
reaches the unity alone, the two dimensional pancake form first.
 In the ZA, the displacement vector is expressed via the initial
 velocity, and for Gaussian initial fluctuations
 92\% of
the Lagrangian particles undergo collapse via pancaking
 (Zel'dovich \& Novikov 1983).
Pancaking is also a typical configuration for the collapsing
homogeneous
 ellipsoid (e.g. Peebles 1980). Since in the first approximation it is
 a reliable generic model in the vicinity of  collapsing region, this
is
a strong argument for the general character of pancaking.
Note that pancakes are pronounced for the baryonic or neutrino
models,
but can have quite complicated ``fractal''-like structure for
the model with power-spectra close to $n=-3$ (Kofman et al 1992).

{\bf Acknowledgements}

We are grateful R.~Bond, S.~Shandarin, A.~Linde and S.~Matarrese for  useful
discussions,
 and S.~Tremaine  for encouraging us to eventually publish the
results,
and for discussions.
We  thank the participants of CITA seminar on February '94, and
University
of Kansas Astronomy seminar on August '93 for useful comments to our
results.

\page

\centerline{\bf APPENDIX}

In this Appendix we write down the explicit expression for the term
containing $ H_{ik} $ in eq.(18) in Newtonian limit (leading terms in
$1/c$ expansion). Using this result we show that eq. (18) and eq.(34)
are
equivalent.
In $1/c$ expansion the time derivative and spatial components of four
-velocity $ u^{\alpha} $ are the quantities of $ c^{-1} $ order. The
spatial part of  projection tensor is then
$ p_{\alpha \beta} = g_{\alpha \beta}.
$ Fixing  indexes $(ik)$ in eq.(18) to be space-like, ($i \to \alpha,
k \to \beta $),
we have
$$
p_{l(\alpha} {\eta _{\beta) r}} ^{sm} u^r {H^l}_{s;m} ~~~
{ \buildrel c \to \infty \over \longrightarrow } ~~~
{1 \over a} \eta _{0 m s (\beta} {H _{\alpha )}}^{s;m} =
{1 \over 2a} \eta _{0 m s (\beta} \eta ^{slnp}
{\left[ C_{\alpha )dnp} u_l u^d \right] } ^{;m}
$$
$$
= -{1 \over a} \left[ {\left( C _{m(\alpha \beta )r} u_0 u^r \right)
}^{;m}+
{\left( C_{m0r(\alpha} u_{\beta)} u^r \right) }^{;m} -
{\left( C_{0(\alpha \beta )r} u_m u^r \right) } ^{;m} \right].
$$
At the last step the identity $ \eta ^{prst} \eta _{iklm} =
-4! \delta _i ^{[p} \delta _k ^r \delta _l ^s \delta _m ^{t]} $ is
used where
brackets denote full antisymmetrization, while parentheses on a pair
of indexes
denote symmetrization. Keeping the leading terms in this formulas, we
get
$$
{\buildrel c  \to \infty \over \longrightarrow } ~~~
{1 \over a} {C_{0 (\alpha \beta ) \gamma} }^{,\gamma}
+ { \left[ C_{\gamma (\alpha \beta) \delta} u^{\delta} \right]
}^{,\gamma}
+ { \left[ C_{0 \gamma 0 (\alpha} u_{\beta)} \right]} ^{,\gamma}
+ { \left[ C_{0 (\alpha \beta ) 0 } u_{\gamma} \right]}^{,\gamma} =
$$
and substituting expressions (25)-(27) for $ C_{iklm} $, we get further
for the last expresion
$$
= {a \over 4} \left[ {h_{0 (\alpha ,\beta) ,\gamma}} ^{,\gamma} -
{h_{0 \gamma ,\alpha ,\beta}} ^{,\gamma} \right] +
{\left[ 2 E_{\gamma (\alpha } u_{\beta )} - 2 E_{\alpha \beta}
u_{\gamma}
- g_{\alpha \beta} E_{\gamma \delta} u^{\delta}
+ g_{\gamma (\alpha } E_{\beta ) \delta} u^{\delta} \right]
}^{,\gamma} .
\eqno(A1)
$$
To proceed further we note that the first term here is equal to
$$ {1 \over c^3 a} \left[ {c^3 \over 2} R_{0 (\alpha ,\beta )} -
\phi ^{\prime}_{,\alpha ,\beta}
- {a^{\prime} \over a} \phi _{,\alpha ,\beta} \right].$$
Next, we use the Einstein equations for $R_{0 \alpha} $, and
decompose
four-velocity
derivative as $ u_{\alpha ,\beta} = \sigma _{\alpha \beta} + {1 \over
3}
\Theta g_{\alpha \beta} $ (assuming irrotational pressureless fluid).
Keeping
in mind the Newtonian expression (28) for $ E_{\alpha \beta} $, it is
straightforward
to obtain finally
$$
p_{l(\alpha} {\eta _{\beta) r}} ^{sm} u^r {H^l}_{s;m} ~~~
{ \buildrel c \to \infty \over \longrightarrow } ~~~
- {1 \over c^3} \left( {1 \over a} \phi ^{\prime} _{,\alpha ,\beta}
+ {a ^{\prime} \over a^2} \phi _{,\alpha ,\beta} \right)
 -{4 \pi G \over c^2} \rho (\sigma _{\alpha \beta} +
{1 \over 3} \Theta g_{\alpha \beta})
-{1 \over 3 c^2} g_{\alpha \beta} \phi ^{,\gamma}_{,\gamma ,\delta}
u^{\delta}
$$
$$
-E_{\alpha \beta ,\gamma} u^{\gamma}
+ 3 E_{\gamma (\alpha } {\sigma ^{\gamma}}_{\beta )} - E_{\alpha
\beta } \Theta
- g_{\alpha \beta} E_{\gamma \delta} \sigma ^{\gamma \delta}  .
\eqno(A2)
$$
Let us now rewrite eq.(18) as
$$
{1 \over c} \left( {1\over a} E^{\prime}_{\alpha \beta}
 + {a^{\prime} \over a^2} E_{\alpha\beta} \right)
+ E_{\alpha \beta ,\gamma} u^{\gamma} +
g_{\alpha \beta} \sigma ^{\gamma \delta} E_{\gamma \delta} + \Theta
E_{\alpha
\beta}
- 3 E_{\gamma (\alpha} {\sigma _{\beta )}}^{\gamma}
+ {4 \pi G \over c^2 } \rho \sigma _{\alpha \beta}
$$
$$
 = - p_{l(\alpha} {\eta _{\beta )r}}^{sm} u^r {H^l}_{s;m} . \eqno (A3)
$$
We see that the magnetic part  $ H_{\alpha \beta} $ from (A2)
 contains largely the same
terms as a left-hand side of equation (A3),  which therefore cancel out.
What is
left takes the form ($ \Theta = {u^{\gamma}}_{,\gamma} $):
$$
-{1 \over 3 c^3} \left[ {\left( g_{\alpha \beta}
{\phi_{,\gamma}}^{,\gamma}
\right) } ^{\prime}
+ {a^{\prime} \over a} g_{\alpha \beta} {\phi _{,\gamma}}^{,\gamma}
\right]
= { 1 \over 3c^2} a \left[ 4 \pi G \rho {u^{\gamma}}_{,\gamma} +
\phi ^{,\gamma}_{,\gamma ,\delta} u^{\delta} \right] g_{\alpha \beta}  .
\eqno(A4)
$$
This expression clearly can be reduced to the derivatives of the
 of Poisson equation:
$$
{1 \over 3} g_{\alpha \beta} \left[
{\partial \over \partial \tau} + {a^{\prime} \over a} +
v^{\gamma} {\partial \over \partial x^{\gamma}} \right]
\left[ \nabla \phi - 4 \pi G a^2 (\rho - \bar \rho) \right] = 0  .
\eqno(A5)
$$

\page

\leftline{\bf REFERENCES}

\def\ref{\par\noindent\hangindent 15pt}

\ref Barnes, A., and Rowlingson, R.  1989,
    Class. Quantum Grav., {\bf 6}, 949.
\ref Bertschinger, E., \& Jain, B. 1994, ApJ, submitted.
\ref Bertschinger, E., \& Gelb, G. 1991, Computers in Physics, Mar/Apr., 164.
\ref Binney, J. \& and Tremaine, S. 1987, {\it Galactic dynamicas}, Princeton
Univ. Press.
\ref Bond, J.R., Klypin, A., Kofman. L. \& Pogosyan, D., 1994. in
preparation.
\ref Bond, J.R., \& Couchman, H., 1988. in {\it Proc. 2-th
Canadian Conf. on GR and Rel. Astrophysics}, eds. Coley A.
\& Dyer C., World Scientific, p. 385.
\ref Cen, R. \& Ostriker, J., 1992, ApJ., {\bf 393}, 22.
\ref Croudace, K., Parry, J., Salopek, D., \& Stewart, J., 1994;
ApJ, {\bf 423}, 22
\ref Ellis, G.F.R., 1971; in {\it General Relativity and Cosmology},
ed. Sachs, R.; NY: Academic Press, p. 104.
\ref Gurbatov, S.N., Saichev, A.I., \& Shandarin, S.F., 1989.
M.N.R.A.S. {\bf 236}, 385;
\ref Hawking, S.W., and Ellis, G.F.R, 1973. {\it The Large Scale
Structure
of Space-Time}, Cambridge University Press.
\ref  Kofman, L.A. \& Shandarin, S.F., 1988. Nature, {\bf 334}, 129.
\ref Kofman, L. 1991, in ``Primordial Nucleosynthesis \& Evolution of
Early
Universe'', eds. Sato, K. \& Audouze, J., Dordrecht: Kluwer.
\ref Kofman, L.A., Melott, A., Pogosyan, D,Yu. \& Shandarin, S.F,
     1992. Apj,  {\bf 393}, 437.
\ref Kofman, L.A., preprint UH-IFA 94/7, astro-ph/9311027
\ref Lachieze-Ray, M. 1993. Apj, {\bf 408}, 403.
\ref Landau, L.D., \& Lifshitz, E.M., 1980, The Field Theory.
\ref Matarrese, S., Pantano, O., \& Saez, D., 1993. Phys. Rev.
 {\bf D47}, 1311.
\ref Matarrese, S., Pantano, O., \& Saez, D., 1994. Phys. Rev.
Lett., {\bf 72}, 320.
\ref  Peebles P.J.E. 1980.
  The Large-scale Structure of the Universe,
     Princeton University Press.
\ref Shandarin, S. \& Klypin, A. 1984. Sov. Astron., 28, 491.
\ref  Shandarin, S.F. and Zel'dovich, Ya.B., 1989. {\it
Rev.Mod.Phys.},
      {\bf 61}, 185.
\ref Tr{\"u}mper, M., 1967. {\it Zeits. f. Astrophys.}, {\bf 66},
215.
\ref  Zel'dovich, Ya.B., 1970. A\&A, {\bf 5} 84.
\ref Zel'dovich, Ya.B. \& Novikov, I.D. 1983, {\it The Structure and
Evolution
of the Universe}, Chicago/London: Univ. Chicago Press.

\end

\bye